\documentclass[a4paper,11pt]{article}
\pdfoutput=1
\usepackage{jheppub}
\usepackage{booktabs}
\usepackage{array}
\usepackage{multirow}
\usepackage{float}
\usepackage{placeins}
\usepackage{lineno}

\newcommand{\miss}{\mathrm{miss}}
\newcommand{\bbbar}{b\bar b}
\newcommand{\ccbar}{c\bar c}
\newcommand{\paperplot}[2][\textwidth]{%
  \IfFileExists{figures/#2}{\includegraphics[width=#1]{figures/#2}}{%
    \fbox{\parbox[c][42mm][c]{0.90\textwidth}{\centering
      Figure pending: \texttt{\detokenize{#2}}}}}}

\title{``Transforming'' LHCb: self-supervised maps of heavy-flavour decays}

\author[a]{Marko Stamenkovic}
\author[a]{and Greg Landsberg}

\affiliation[a]{Department of Physics, Brown University,\\
Providence, Rhode Island 02912, U.S.A.}
\emailAdd{marko.stamenkovic@cern.ch}

\abstract{%
Decays of beauty and charm hadrons provide sensitive probes of physics beyond the standard model, including decays with invisible particles, in which part of the final state leaves no reconstructed detector signature. The large heavy-flavour data samples recorded by the LHCb experiment at the CERN LHC, together with its precise tracking, displaced vertex reconstruction, and particle identification, make it particularly well suited to learning a map of reconstructed heavy-hadron decay environments directly from data. We propose to bring the recent advances in jet flavour tagging in ATLAS and CMS to significantly improve on the performance of the current LHCb taggers and extend them to the reconstruction of heavy-flavour decays with several invisible particles in the final state. To achieve this, we introduce a self-supervised transformer architecture that learns the decay maps without flavour or exclusive-decay labels by inferring masked particle identification information and completing jets from which constituents have been removed. Across five classification tasks in simulated LHCb Open Data, the self-supervised model outperforms an otherwise identical transformer with random weights, and performs comparably to a fully supervised transformer model. We achieve a tagging power of approximately 10\%. In addition, removing constituents from reconstructed exclusive decays also systematically increases the model anomaly score relative to random removals from the same heavy hadrons. We confirm this behaviour directly in 2017 LHCb proton-proton collision Open Data: the score increases for all eight studied heavy-flavour channels, and the signal region response exceeds that in the adjacent sidebands. These studies provide a proof of principle that mapping heavy-flavour decay environments through jets can transform flavour tagging in LHCb and extend the discovery reach for incomplete or otherwise unusual decays.}


\begin{document}
\maketitle
\flushbottom

\section{Introduction}
\label{sec:introduction}

Heavy-flavour decays provide sensitive probes of physics beyond the standard model (SM). The weak heavy-flavour decays can reveal new sources of flavour and $CP$ violation, while light invisible particles can produce final states that are only partially reconstructed~\cite{Blake:2016olu,He:2023kec}. A particularly compelling example is $B$-mesogenesis, in which $CP$-violating neutral $B$ meson oscillations and subsequent decays generate both the visible matter-antimatter asymmetry and the dark matter relic abundance~\cite{Elor:2019hpa}. In this framework dark matter carries baryon number, so an asymmetry in visible baryons is balanced by an opposite asymmetry in the dark sector. Characteristic collider signatures include exotic $B$ meson decays to a visible baryon and missing energy. Searches for ``incomplete'' heavy-flavour decays can therefore connect collider measurements to the two central cosmological questions: the origin of the asymmetry between matter and antimatter and the nature of dark matter.

The CERN LHC produces beauty and charm hadrons in enormous numbers, and the LHCb experiment~\cite{LHCb:2008vvz,LHCb:2014set} is built to turn this high-volume data into detailed decay information for each recorded event. The forward geometry precise tracking and vertexing resolve displaced heavy-flavour hadron decays, while the ring-imaging Cherenkov (RICH) detectors identify charged hadrons over a broad momentum range~\cite{LHCb:2014set,Adinolfi:2012qfa}. Periodic reversal of the dipole magnetic field exchanges the bending direction of positive and negative particles and provides an important handle on detector effects that depend on the charge. The combination of large heavy-flavour samples, precise reconstruction and particle identification makes LHCb particularly well suited to mapping heavy-flavour hadron decay environments through their visible constituents.

The state-of-the-art flavour tagging in LHCb~\cite{LHCb:2025okz} is currently focused on secondary vertices and underutilizes the important information about the full fragmentation history, the underlying event, and the pileup. Furthermore, it is based on the \textsc{DeepSets} deep neural network (DNN) architecture~\cite{Zaheer:2017wmg}, which is known to be less powerful for tagging than more modern graph neural network or transformer architectures. The advances in jet flavour tagging in the ATLAS~\cite{ATLAS:2025nyf,ATLAS:2025dkv} and CMS~\cite{CMS-DP-2025-081,CMS:2026vpb} experiments at the LHC of the past few years have demonstrated that the representation of a jet as a permutation-invariant particle cloud allows optimal performance of jet tagging algorithms to be achieved, offering an order of magnitude better background rejection compared to older taggers that were not using the entire information within a heavy-flavour jet. For example, the state-of-the-art transformer-based tagger used for the recent CMS $B^0_{(s)} \to J/\psi K_S^0$ time-dependent $CP$-violation analysis achieved an impressive tagging power of 6--6.5\% for the same-side tagging by utilizing tracks within a large-radius jet centered on the decay of interest~\cite{CMS-PAS-BPH-26-005}. This demonstrates that using {\it non-local\/} information in the $\eta$-$\phi$ space (i.e., particles that are separated by relatively large $\Delta R = \sqrt{(\Delta\eta)^2 + (\Delta\phi)^2}$ values) is useful in the transformed hyperspace used for tagging, as some of these particles are, in fact, adjacent to the decay products of interest in this new {\it tagging\/} space.

We use these novel tagging architectures to introduce the mapping of heavy-flavour decays through their visible constituents using self-supervised particle set completion. Each heavy-flavour hadron decay environment is represented by its reconstructed particles and the detector information. During pretraining, the PID inputs are masked and up to four decay constituents are removed completely. All quantities that depend on the surviving set are rebuilt, and a transformer predicts the multiplicity, kinematics, charge, identity, and topology of the missing content. The transformer condenses the surviving decay environment into a continuous global representation---the learned hadronization and decay map---in which environments requiring similar completions can acquire similar coordinates. The magnet polarity is provided to the model as an additional input feature.

The study follows two stages. First, simulated LHCb Open Data provide controlled tests: generator-level flavour information can be used only for downstream probes, and reconstructed charm and beauty decays can be removed and compared with matched tracks from the same heavy-flavour decay environment. Second, the architecture is trained anew on LHCb Open Data of 2017 proton-proton collision data, without flavour or exclusive decay labels, and validated afterward on eight reconstructed charm and beauty modes. The collision data study tests whether the same learned notion of completeness is visible in genuine reconstructed decays in the collision data. 

This work builds on particle set and attention-based jet models~\cite{Komiske:2018cqr,Qu:2019gqs,Qu:2022mxj}, self-supervised and masked particle learning~\cite{Dillon:2021gag,Golling:2024nki,Leigh:2024pqa}, and recent transferable jet representations~\cite{Birk:2024knn,Letellier:2026jpjepa}. It is tested using public simulated LHCb jets, whose tracking and particle identification (PID) information make reconstructed heavy-flavour structure accessible~\cite{LHCb:2014set,Adinolfi:2012qfa,LHCb:2015nta}. The same sample has supported earlier studies of flavour and charge tagging~\cite{Felser:2020cqc,Gianelle:2022fda}. The construction is also conceptually related to partial decay reconstruction~\cite{Freytsis:2014hpa}.

The original step in this approach is to delete complete reconstructed constituents rather than only hiding their features, and then to validate the learned notion of completeness through removal tests performed for individual candidates. Candidate tracks are selected independently of the model response and are compared with matched track groups from the same jets. This connects pretraining with missing constituents directly to reconstructed heavy-flavour objects, while avoiding an exclusive label during representation learning. Finally, we perform an exploratory anomaly search in collision data, using the learned map to organize high-missingness jets by similarity and identify recurrent unusual reconstructed topologies.

This paper is organized as follows. Section~\ref{sec:data} describes the simulated LHCb Open Data sample and its event-level partition. Section~\ref{sec:method} presents the self-supervised mapping task and model. Section~\ref{sec:validation} demonstrates controlled completion, flavour structure, and reconstructed decay interventions in simulation. Section~\ref{sec:collision-data} demonstrates the power of the method using LHCb Open Data 2017 proton-proton collisions. Sections~\ref{sec:discussion} and~\ref{sec:conclusion} discuss the scope and summarize the results.

\section{Simulated LHCb Open Data sample}
\label{sec:data}

\subsection{Open data sample}

We use the LHCb Open Dataset of simulated dijet events from proton-proton collisions~\cite{LHCbOpenData:2020jets}, produced at a collision energy of $13$~TeV. The jets are reclustered with the anti-$k_{\rm T}$ algorithm using a distance parameter $R=0.5$~\cite{Gianelle:2022fda}. The proton interactions and jet fragmentation are generated with \textsc{Pythia}~8~\cite{Sjostrand:2014zea}, while beauty hadron decays are modelled with \textsc{EvtGen}~\cite{Lange:2001uf}. We restrict this proof-of-principle study to $R = 0.5$ jets for practical simplicity, since the available public simulation is provided in this format. For example, extending these studies to jets of a larger distance parameter, e.g. $R = 0.8$, often used in ATLAS and CMS, may indicate further improvement of the performance when an even larger set of particles in the vicinity of the decay of interest is used.

The detector response is simulated with \textsc{Geant4}~\cite{GEANT4:2002zbu} and reconstructed with the LHCb software framework~\cite{Felser:2020cqc,LHCb:2014set}. The sample includes inclusive $\bbbar$, $\ccbar$ and light-parton dijets, together with $Z\to\bbbar$ events, over several intervals of jet transverse momentum and both magnet polarities. Here $\bbbar$ denotes a beauty quark-antiquark pair, while $\ccbar$ denotes a charm quark-antiquark pair. When the quark charge sign is not distinguished, a $b$-jet denotes a jet matched to either a $b$ or $\bar b$ parton, and a $c$-jet denotes a jet matched to either a $c$ or $\bar c$ parton. The composition is summarized in Table~\ref{tab:dataset-composition}.

Each simulated event contains two reconstructed jets, each selected to contain a reconstructed secondary vertex. The stored constituents include kinematic and displacement information together with probabilistic RICH PID responses~\cite{Adinolfi:2012qfa}. Generator-matched parton flavour is used only for diagnostics: no constituent ancestry or exclusive heavy-flavour hadron decay generator-level information is available.

\begin{table}[H]
\centering
\small
\begin{tabular}{lrrr}
\toprule
Source category & Events & Jets & Fraction of events \\
\midrule
$\bbbar$ dijets       & 605,776 & 1,211,552 & 82.38\% \\
$\ccbar$ dijets       &  77,357 &   154,714 & 10.52\% \\
Light parton dijets   &   8,604 &    17,208 &  1.17\% \\
$Z\to\bbbar$          &  43,611 &    87,222 &  5.93\% \\
\midrule
Total                 & 735,348 & 1,470,696 & 100\% \\
\bottomrule
\end{tabular}
\caption{Composition of the simulated sample. The source categories are used
only for bookkeeping and downstream validation.}
\label{tab:dataset-composition}
\end{table}

\subsection{Inputs and event-level partition}

Each jet is represented as an unordered set of at most 64 reconstructed constituents. The complete input representation is listed in Table~\ref{tab:inputs}. Variables spanning broad positive ranges are logarithmically transformed to reduce their dynamic range before standardization. The RICH and calorimeter PID response variables are input features. They are hidden for particles selected by the masked PID task. The reconstructed PID class is never an input. It provides a target with 15 classes,
\begin{equation*}
 (e^-,e^+,\mu^-,\mu^+,\gamma,\pi^0,\pi^+,\pi^-,K^+,K^-,p,\bar p,
 K^0_{\rm S},\Lambda,\bar\Lambda).
\end{equation*}

\begin{table}[H]
\centering
\small
\begin{tabular}{@{}lp{0.72\textwidth}@{}}
\toprule
Input group & Reconstructed inputs \\
\midrule
Particle kinematics & $\log p_{\mathrm T}$, $\log p$, momentum fractions relative to the jet, $\Delta\eta$ and $\Delta\phi$ relative to the jet axis, and electric charge. \\
Track and displacement & Signed $\log(1+\mathrm{IP})$, raw signed IP, $\log(1+\mathrm{IP}\chi^2)$, $\log(1+\chi^2_{\rm trk})$, track state position relative to the primary vertex, track direction, and availability flags. \\
PID & Five RICH neural network responses, $\mathrm{ProbNN}(e,K,p,\pi,\mu)$, and four calorimeter responses ($E_{\rm ECAL}$, $E_{\rm HCAL}/E_{\rm ECAL}$, $E_{49}$ and PRS), each with its availability flag. \\
Pairwise relations & $\log\Delta R$, pair mass proxy under a common pion hypothesis, charge product, proxy for the distance of closest approach, and its availability. \\
Jet context & $\log p_{\mathrm T}^{\rm jet}$, jet $\eta$, massless jet mass, constituent multiplicity, charged fraction, jet width, primary vertex multiplicity, displaced track multiplicity, vertex counts for two, three and four tracks, cascade count, and vertex/cascade truncation flags. \\
Conditioning & Magnet polarity. \\
\bottomrule
\end{tabular}
\caption{Reconstructed inputs to the transformer.}
\label{tab:inputs}
\end{table}

The dataset is partitioned at the event level, so both jets from the same event always belong to the same partition. Table~\ref{tab:splits} gives the 65\%, 15\%, 10\% and 10\% training, validation, test and analysis split. The validation sample determines when the model training is stopped. The test sample supplies the final completion and flavour probe metrics. The analysis sample is reserved for reconstructed decay interventions.

\begin{table}[H]
\centering
\begin{tabular}{lrrrl}
\toprule
Partition & Fraction & Events & Jets & Use \\
\midrule
Training   & 65\% & 477,871 & 955,742 & self-supervised optimization \\
Validation & 15\% & 110,331 & 220,662 & checkpoint and probe selection \\
Test       & 10\% &  73,639 & 147,278 & final frozen metrics \\
Analysis   & 10\% &  73,507 & 147,014 & reconstructed decay interventions \\
\bottomrule
\end{tabular}
\caption{Data partitions with disjoint events. The analysis sample does not enter
training, early stopping, class weighting or probe selection.}
\label{tab:splits}
\end{table}

\section{Self-supervised mapping of hadronization and decay environments}
\label{sec:method}

The core of this proof-of-principle is to exploit the precise reconstruction of LHCb to map the hadronization and decay of heavy-flavour hadrons. A self-supervised transformer is trained to infer masked PID information and particles removed from the reconstructed decay environment, thereby learning a representation of each heavy-flavour hadron. To demonstrate that this representation captures physics, we first test flavour and quark charge sign classification, and then measure the model response to controlled particle removals from reconstructed charm and beauty decays.

\subsection{Deletion and reconstruction of a jet view}

During the initial self-supervised training used to construct the hadronization and decay map, each complete jet is paired with an artificial view in which zero to four reconstructed constituents are removed. The probabilities for removing zero, one, two, three or four constituents are chosen to be 0.25, 0.35, 0.25, 0.10 and 0.05, respectively. The $P=0$ case is an unchanged reference jet. For $P>0$, one of six artificial removal types selects the constituents to be deleted. Table~\ref{tab:removal-types} defines these types and their proposal weights.

\begin{table}[H]
\centering
\small
\begin{tabular}{@{}lp{0.14\textwidth}p{0.56\textwidth}@{}}
\toprule
Removal type & Proposal weight & Selection rule \\
\midrule
Random charged & 0.25 & Delete $P$ charged-particle tracks with valid reconstructed PID and track state information. \\
Random neutral & 0.10 & Delete $P$ neutral constituents with valid reconstructed PID. \\
Opposite sign pair & 0.20 & For $P=2$ only, delete one positively and one negatively charged track. \\
Angularly local group & 0.15 & Select an eligible constituent and delete it with its $P-1$ nearest eligible neighbours in the $(\eta,\phi)$ plane. \\
Displaced vertex & 0.18 & Select one eligible reconstructed displaced vertex and delete $P$ of its member tracks. \\
Cascade & 0.12 & Select one eligible reconstructed cascade decay and delete $P$ of its member tracks. \\
\bottomrule
\end{tabular}
\caption{Artificial removal types for $P>0$. Proposal weights are renormalized
over the types that are feasible for the selected jet and value of $P$.}
\label{tab:removal-types}
\end{table}

For the vertex and cascade removal categories, tracks are grouped using only their reconstructed trajectories, without PID information. A displaced vertex is built from two, three or four charged tracks with compatible straight line trajectories and a displaced common point. A cascade combines such a vertex with an additional bachelor track that forms a second, pointing vertex. These are geometric grouping tools, not exclusive decay reconstructions or truth labels. The removal selection is deterministic for each jet within a training epoch and is redrawn at every epoch.

Removed constituents are deleted entirely from the model input, and the surviving particles are compacted so that no placeholder identifies a removed particle or its former position. To avoid biases, we apply the same reconstruction to the complete view and to views with one, two, three or four removed particles. The jet quantities are therefore rebuilt only from the particles present in each view.

\subsection{Transformer and prediction heads}

The encoder contains five transformer blocks with normalization before each block, embedding dimension 96, eight attention heads, feed-forward dimension 384, and dropout of 0.1. No absolute positional encoding is used. Pairwise reconstructed relations enter as attention biases. Alongside the particle representations, the encoder maintains a learned global {\it completeness\/} representation. This representation does not correspond to a reconstructed object: it is initialized from a trainable 96-dimensional vector and combines information from all surviving particles through the attention mechanism to form a global summary of the residual jet. All completion predictions, including the missing particle count used for the missingness score, are made from this final summary. A decoder with one layer maps the summary to four exchangeable missing particle queries. The complete model has approximately one million trainable parameters.

The attention heads predict:
\begin{enumerate}
  \item the number of missing particles $P=0,\ldots,4$;
  \item one of the six corruption mechanisms;
  \item eleven combined properties of the removed set, including energy,
        momentum components relative to the residual axis, mass, charge, and counts;
  \item the properties of up to four removed particles, including six
        kinematic and displacement features, charge and signed reconstructed PID; and
  \item the masked PID class and nine binned PID responses for a separate sample
        of surviving particles whose PID inputs are masked while their other
        measured properties remain available to the encoder.
\end{enumerate}
For the last objective, the model predicts the PID of surviving particles after their PID inputs have been hidden. It is trained at the same time as the completion task for deleted particles.

\subsection{Loss, balancing, and optimization}

The total loss function is given by
\begin{align}
\mathcal L ={}&
0.10\mathcal L_{\rm PID}^{\rm nat}
+0.10\mathcal L_{\rm PID,species}^{\rm bal}
+1.00\mathcal L_{P}^{\rm nat}
+0.20\mathcal L_{\rm set} \nonumber\\
&+0.15\mathcal L_{\rm summary}
+0.15\mathcal L_{\rm topology}
+0.05\mathcal L_{\rm type}
+1.00\mathcal L_{\rm missPID}^{\rm bal}.
\label{eq:total-loss}
\end{align}
We choose the relative loss weights before training and keep them fixed throughout. An optimization of these weights is beyond the scope of this proof-of-principle study. The terms are:
\begin{itemize}
  \item $\mathcal L_{\rm PID}^{\rm nat}$: prediction of the masked PID class
        and PID response bins of surviving particles, weighted according to their
        natural frequency in the training sample.
  \item $\mathcal L_{\rm PID,species}^{\rm bal}$: an additional masked PID
        class loss in which rare reconstructed particle hypotheses receive larger
        weights.
  \item $\mathcal L_{P}^{\rm nat}$: classification of the number $P=0,\ldots,4$
        of particles removed from the jet, using the natural training frequency.
  \item $\mathcal L_{\rm set}$: the equally weighted existence, continuous property
        and charge losses for the matched individual missing particle slots
        described above.
  \item $\mathcal L_{\rm summary}$: prediction of combined properties of the
        entire removed set, such as its total momentum and charge.
  \item $\mathcal L_{\rm topology}$: binary classification of the topology labels
        associated with each removed particle.
  \item $\mathcal L_{\rm type}$: classification of the artificial removal mechanism
        used to construct the incomplete jet view.
  \item $\mathcal L_{\rm missPID}^{\rm bal}$: prediction, balanced across classes, 
        of the reconstructed PID of each removed particle.
\end{itemize}
$\mathcal L_{P}^{\rm nat}$ and $\mathcal L_{\rm set}$ constrain the missing particle multiplicity independently. The former compares the predicted count with the true value of $P$, while the existence term in $\mathcal L_{\rm set}$ compares the individual particle predictions with the true removed set. Neither prediction is passed to the other. If two particles are removed but both heads
predict only one, both losses are penalized.

All parameters are initialized randomly. Training uses the \textsc{AdamW} optimizer with the learning rate of $6\times10^{-4}$, weight decay $10^{-2}$, 5\% warm-up followed by cosine decay, clipping of the gradient norm at one, and a global batch size of 1024 on eight GPUs. For the completion task, the model uses the available PID responses of the surviving particles. At the same time, a second view of each jet is created in which the PID inputs of 15\% of the surviving particles are hidden. The model predicts their reconstructed particle class and nine binned PID responses. Early stopping selects epoch 60 and stops after epoch 72 with a patience of 12. The training run takes 4.16 hours on eight RTX A5000 GPUs, approximately 205 seconds per epoch.

\subsection{Missingness score}
\label{sec:score}

For an input jet $X$, the natural count head returns probabilities for each possible number of removed particles, $P=0,\ldots,4$. We denote the probability that no particle was removed by $p_0(X)=p_\theta(P=0\mid X)$. The artificial removal uses an unchanged jet in 25\% of cases, so $\pi_0=0.25$ is the baseline
fraction of original jets.

We define
\begin{equation}
 S_{\miss}(X) =
 \log\frac{1-p_0(X)}{p_0(X)}
 -\log\frac{1-\pi_0}{\pi_0}.
 \label{eq:missingness-score}
\end{equation}
The first term is the log probability of the model that the jet is incomplete rather than complete. The second term is the corresponding log probability in the training sample before looking at any jet features. Their difference therefore compares the model prediction with the artificial removal baseline. A model that can do no more than return the training fraction has $p_0(X)=\pi_0$ and hence $S_{\miss}(X)=0$. Positive scores indicate that the reconstructed jet is more compatible with the artificial incomplete training population than with a complete jet. The prior subtraction is the same constant for every jet, so it cancels in the score differences used in the track removal tests. The score is constructed from the model probability that at least one constituent is missing, inclusive over all six corruption mechanisms and removal multiplicities $P=1,\ldots,4$.

\section{Demonstration in simulation}
\label{sec:validation}

We test the learned map of heavy-flavour hadronization and decay at three complementary levels. Controlled constituent deletion measures whether the model solves the task on unseen events by comparing its response to pairs of complete and corrupted jets. Probes trained with limited labels quantify how much heavy-flavour and quark charge sign information is retained in the frozen representation. Finally, we construct artificial anomalies by removing the constituents of resonances in reconstructed exclusive decays and test whether the model recognizes the resulting missing decay structure.

\subsection{Controlled constituent deletion}

Each jet in the test set is either left unchanged or receives one artificial removal. The test sample contains approximately 37,000 unchanged jets with $P=0$ and 110,000 corrupted jets with $P>0$. For every jet with $P>0$, we construct both the residual view, with its selected constituents removed, and a clean view reconstructed from the same original jet without a deletion. We run the frozen model on both views and calculate the paired score shift
\begin{equation}
 \Delta S_{\rm test}=S_{\miss}(\text{residual view})-S_{\miss}(\text{complete view}).
 \label{eq:test-score-shift}
\end{equation}
The paired comparison therefore contains approximately 110,000 controlled tests before and after removal.

\begin{figure}[h]
\centering
\paperplot{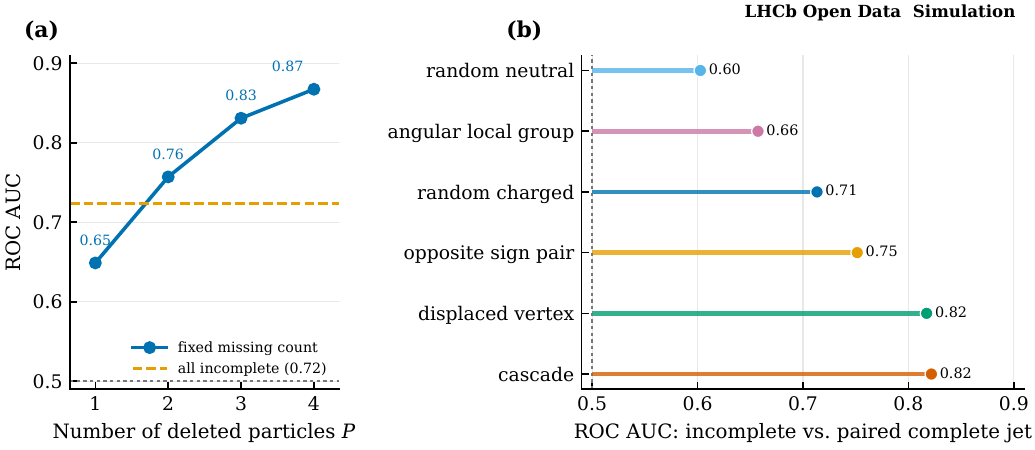}
\caption{Performance for artificial deletions in the independent test sample. The left panel shows the paired discrimination between complete and incomplete views as a function of the number of removed constituents. The right panel resolves the AUC by deletion mechanism.}
\label{fig:synthetic}
\end{figure}

We quantify how well the missingness score separates the incomplete views from their paired complete views with a receiver operating characteristic (ROC) curve. The ROC curve scans a score threshold, and its area under the curve (AUC) is 0.50 for no separation and 1.00 for perfect separation. The inclusive AUC is 0.72. As shown in Figure~\ref{fig:synthetic}, the response rises monotonically with the number of removed constituents. The AUC is 0.65, 0.76, 0.83 and 0.87 for $P=1,2,3,4$, respectively. The corresponding mean paired score shifts are 0.79, 1.71, 2.58 and 3.19.

Under the realized mechanism and multiplicity mixtures, the largest
AUCs are 0.82 for tracks drawn from a displaced vertex and 0.82 for a cascade decay removal, compared with 0.71 for a random charged particle, 0.60 for a neutral particle, 0.75 for an opposite-sign track pair, and 0.66 for an angularly local group. Because feasible mechanisms have different $P$ distributions, these values do not isolate topology at a fixed multiplicity.

Exact missing multiplicity is a more difficult task than binary missingness. The natural count head obtains 39.6\% accuracy for the natural class mixture and 27.7\% when the five classes receive equal weight. Charge is predicted with 86.2\% accuracy, or 83.0\% when its classes receive equal weight. The missing PID head obtains 30.7\% accuracy for the natural class mixture and 46.1\% when the PID classes receive equal weight. These diagnostics are reported in Appendix~\ref{app:training}.

\subsection{Flavour and quark charge sign tagging with limited labels}
\label{sec:flavour-probes}

To demonstrate that the learned map encodes relevant physics about heavy-flavour decays, we test whether it retains information about the flavour and charge of the initiating quark, i.e. classical jet flavour tagging. Neither flavour nor charge information is used during self-supervised training. Each complete $P=0$ jet is passed through the encoder and summarized by a 96-dimensional embedding vector $z$. For each binary task, the only object fitted to generated-level parton labels is an affine probe, $\alpha z+\beta$, with two output scores and no hidden layers. The probe parameters $\alpha$ and $\beta$ are optimized with cross entropy weighted by class using the \textsc{AdamW} optimizer. The epoch is selected using the labelled validation split. The transformer weights are fixed throughout, so gradients from this probe never update the encoder.

The self-supervised probe uses the encoder trained jointly on the particle completion and masked PID tasks. For validation, we use the same transformer architecture with randomly initialized, frozen weights. Both probes are therefore equally simple linear readouts of a fixed representation with 96 dimensions. If the self-supervised probe outperforms the random weight transformer, the difference is information made accessible by the joint self-supervised training rather than a separation that can be obtained from the raw inputs through the architecture alone. Five conditional binary tasks are considered: $b$ vs. $c$, $b$ vs. light, $c$ vs. light, $b$ vs. $\bar b$, and $c$ vs. $\bar c$. 

\begin{figure}[h]
\centering
\paperplot[0.98\textwidth]{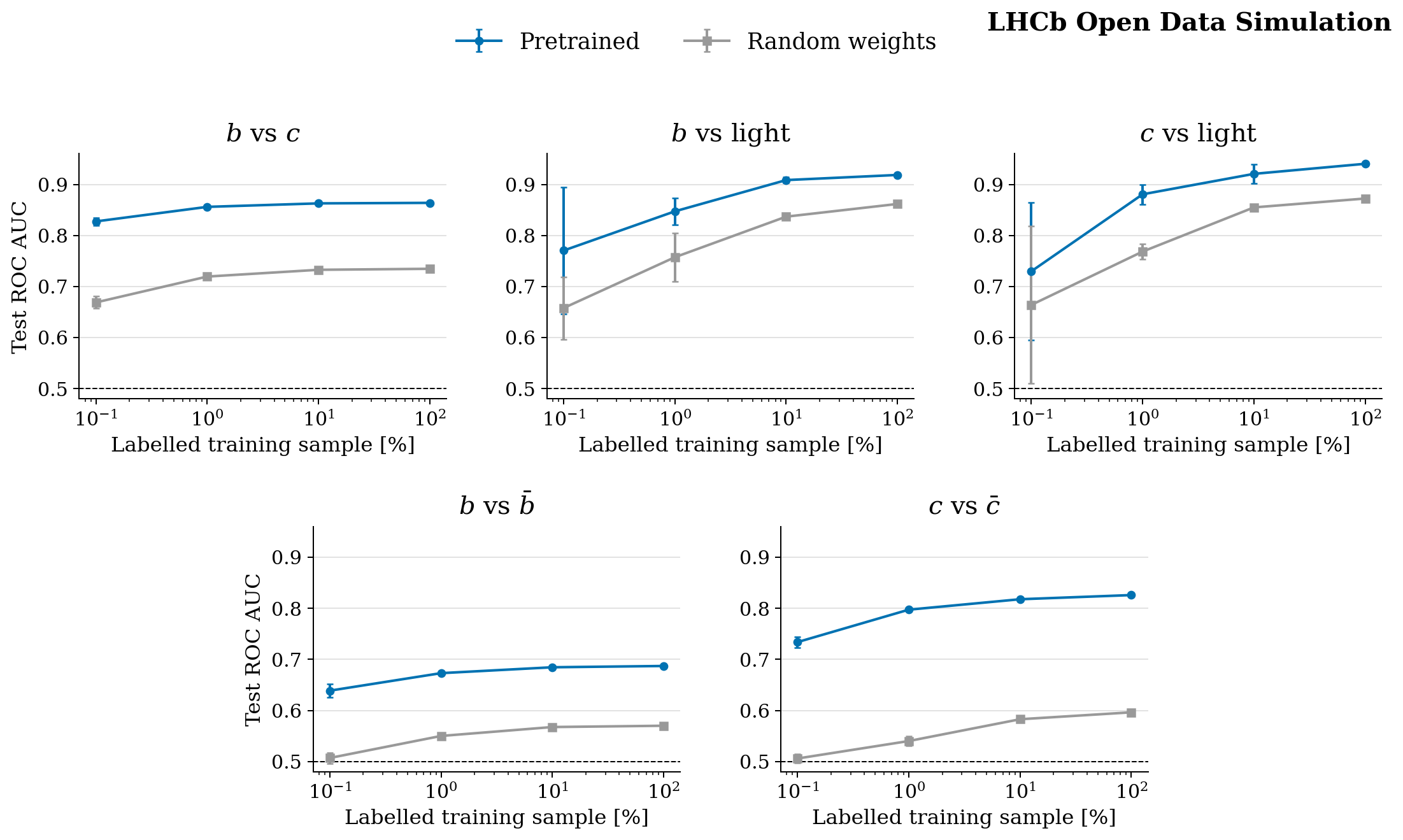}
\caption{Test ROC AUC of affine probes trained with 0.1\%, 1\%, 10\%, and 100\%
of the labelled training events. Points and error bars show the mean and standard
deviation over five subsets for the frozen pretrained and random weight encoders.}
\label{fig:label-efficiency}
\end{figure}

To quantify label efficiency, we repeat the probe fit with training subsets containing $0.1\%$, $1\%$, $10\%$ and $100\%$ of the labelled events. At each fraction, the pretrained and random encoders use identical subsets, while the validation and test sets remain fixed. Figure~\ref{fig:label-efficiency} shows the mean and standard deviation from the five subsets. The pretrained representation outperforms random features for every task and label fraction. For $b$ versus $c$, it reaches an AUC of 0.83 with only $0.1\%$ of the labelled training events, compared with 0.67 for random features, and reaches 0.86 with $1\%$ of the labelled training events. The broader bands for the heavy vs. light tasks reflect the small test sample of light jets. The consistent performance improvement over the frozen random encoder shows that the self-supervised training captures correlations among multiple particles associated with heavy hadron decay environments. This demonstrates that the learned map encodes physically meaningful flavour and quark charge sign information. The ability to apply a common learned representation across distinct problems and recover much of the performance using only a small fraction of labelled data are defining qualities of foundation models. These results provide an initial indication of such behaviour in heavy-flavour physics.

\FloatBarrier

\begin{figure}[tbp]
\centering
\paperplot{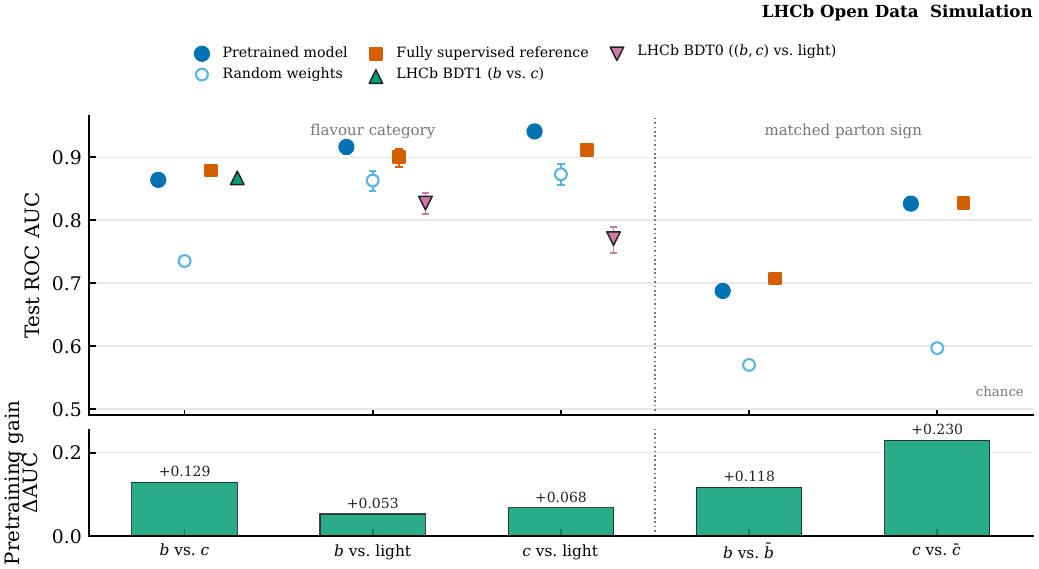}
\caption{Test ROC AUC for five flavour and quark charge sign classification tasks.}
\label{fig:flavour}
\end{figure}

Having established that the frozen representation retains flavour information, we next test how much of the performance available to a model trained end to end with all labels can be recovered by fitting only the affine probe. We train the same transformer architecture in a single supervised training over the five classification tasks, using the full labelled training sample. This provides a reference in which the labels shape the entire representation rather than only the affine probe. Figure~\ref{fig:flavour} compares three controls on the same jets in the test sample. First, the frozen random encoder tests whether the probe can obtain separation from the untrained architecture alone. Second, the fully supervised reference provides the comparison to an architecture matched model trained end to end. Third, the stored LHCb flavour tagging BDT outputs are evaluated directly, with BDT1 used for $b$ versus $c$ and BDT0 for the two tasks that separate heavy-flavour jets from the light-parton ones~\cite{LHCb:2015nta}.

With all labelled training events available, the affine probes recover 97--100\% of the compact fully supervised performance across the five tasks. The nominally better performance of the pretrained probes for the light-jet tasks should not be overinterpreted because the simulated light jet sample is small. With larger samples, a fully supervised training is expected to set the performance ceiling for this type of study. Relative to the stored LHCb taggers, the pretrained probe recovers 99.6\% of the BDT1 AUC for $b$ versus $c$ and improves on BDT0 by 11\% and 22\% for $b$ versus light and $c$ versus light, respectively. The agreement with the supervised and legacy references reinforces the conclusion from Figure~\ref{fig:label-efficiency}: the self-supervised training retains physically useful heavy-flavour structure.

\subsubsection{Dilution and tagging power}

For comparison with flavour taggers used in data, we adopt the event-dependent
definition used by LHCb,
\begin{equation}
 D_i = 1-2\omega_i,
 \qquad
 P_{\rm tag}=\epsilon_{\rm tag}\left\langle D_i^2\right\rangle,
 \label{eq:tagging-power}
\end{equation}
where $\epsilon_{\rm tag}$ is the fraction assigned a tag and $\omega_i$ is the mistag probability for jet or candidate $i$. The probe assigns $b$ or $\bar b$ according to whether its probability is above or below $0.5$, and $\omega_i$ is the probability assigned to the opposite decision. Every selected jet receives a prediction, giving a conditional tagging efficiency of $100\%$.

We use each jet in a simulated $b\bar b$ event to infer the sign of the other jet. Within the stored dijet sample, this gives $D_{\rm eff}=\sqrt{\langle D_i^2\rangle}=32.6\%$ and $P_{\rm tag}=10.6\%$. We apply the same procedure to simulated $c\bar c$ events and obtain $D_{\rm eff}=55.8\%$ and $P_{\rm tag}=31.1\%$, substantially higher than for the beauty sample.

\begin{table}[H]
\centering
\caption{Opposite jet tagging performance in the simulated $\bbbar$ and $\ccbar$ samples, with published LHCb inclusive beauty flavour tagging results for comparison. The effective dilution is $D_{\rm eff}=\sqrt{\langle D_i^2\rangle}$.}
\label{tab:tagging-power-comparison}
\small
\setlength{\tabcolsep}{4pt}
\begin{tabular}{p{0.27\textwidth} p{0.29\textwidth} >{\centering\arraybackslash}p{0.11\textwidth} >{\centering\arraybackslash}p{0.11\textwidth} >{\centering\arraybackslash}p{0.11\textwidth}}
\toprule
Method & Evaluation sample and task & $\epsilon_{\rm tag}$ [\%] & $D_{\rm eff}$ [\%] & $P_{\rm tag}$ [\%] \\
\midrule
This work, opposite jet
  & Inclusive $b$-jets
  & $100$ & $32.6$ & $10.6$ \\
LHCb inclusive tagger~\cite{LHCb:2025okz}
  & Exclusive $B^0$ decays
  & $95.0$--$98.5$ & $23.7$--$24.8$ & $5.4$--$5.9$ \\
LHCb inclusive tagger~\cite{LHCb:2025okz}
  & Exclusive $B_s^0$ decays
  & $88.0$--$96.9$ & $24.0$--$29.7$ & $5.4$--$7.8$ \\
\midrule
This work, opposite jet
  & Inclusive $c$-jets
  & $100$ & $55.8$ & $31.1$ \\
\bottomrule
\end{tabular}
\end{table}

The comparison with LHCb is conceptually close. In the LHCb measurements, an exclusive $B$ candidate is reconstructed and its production flavour is inferred from the remaining particles in the event. Here, one jet in a simulated $b\bar b$ event is used to infer the sign of the other jet. The important difference is the sample selection. Our sample already contains two reconstructed jets with secondary vertices, so the quoted $100\%$ efficiency is conditional on the availability of an accepted opposite side $b$-jet. In data, a $b$-jet selection would be needed to reject $c$-jets and light jets, and candidates without an accepted opposite-side $b$-jet would be untagged. A direct candidate-level comparison would therefore need to include this selection efficiency and calibrate the mistag probability in data. Moreover, our inclusive samples contain charged heavy hadrons such as $B^\pm$ and $D^\pm$, whose charge makes the quark charge sign easier to identify than for neutral $B^0$ and $B_s^0$ candidates. The comparison should therefore not be interpreted as an improvement over the LHCb tagger. Instead, this approach is complementary and could extend quark charge sign identification to inclusive heavy-flavour decays beyond the exclusive modes used by the existing methods.

\FloatBarrier

\subsection{Reconstructed decay interventions}
\label{sec:reco-simulation}

We test whether the learned representation responds to recognizable decay structures by removing reconstructed charm resonances and topologies enriched in beauty decays. Their score shifts are compared with mass sidebands and matched tracks from the same jet.

\subsubsection{Control samples and intervention}

\begin{figure}[tbp]
\centering
\paperplot{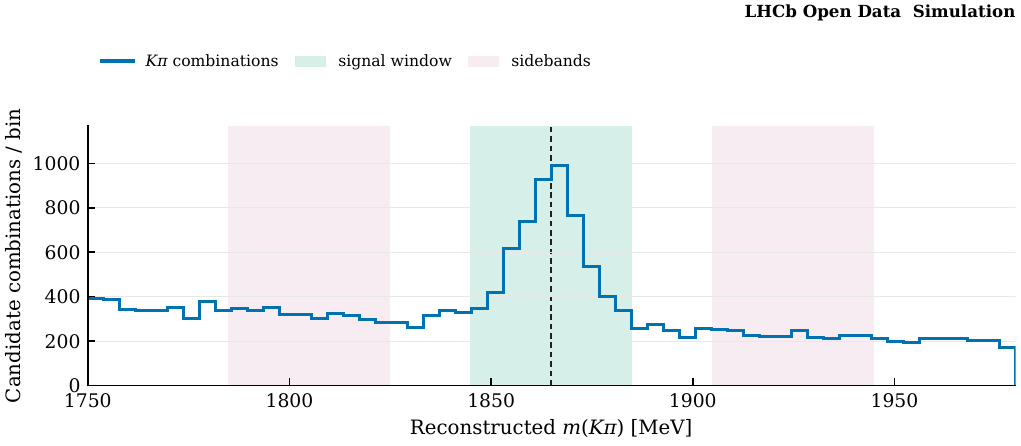}
\caption{Reconstructed $K\pi$ mass spectrum for right-sign candidates in $b$-jets. The signal window around the $D^0$ mass and the two sidebands used in the primary track removal test are shaded. Charge-conjugate decays are combined.}
\label{fig:d0-mass}
\end{figure}

Candidate construction is performed only in the analysis split and is independent of the model response. It uses reconstructed charge, momentum, PID responses, track reference positions and displacement variables. No missingness output enters candidate selection. World-average particle masses are taken from the Particle Data Group review~\cite{ParticleDataGroup:2024cfk}. The principal categories are:
\begin{itemize}
  \item two track $D^0/\bar D^0\to K^\mp\pi^\pm$ candidates in $b$-jets.
  \item $\phi\to K^+K^-$ in $b$-jets.
  \item three track $D^0\pi$ and four track $D^\pm\pi$ topologies in $b$-jets,
        enriched in $B^\pm$ and $B^0/\bar B^0$ decays, respectively.
  \item two track right sign $D^0/\bar D^0\to K^\mp\pi^\pm$ candidates in
        $c$-jets matched to partons with known sign.
  \item $D^{*\pm}\to D^0(\bar D^0)\pi^\pm$ using the $\Delta m$ peak in $c$-jets.
  \item $D^\pm\to K^\mp\pi^\pm\pi^\pm$ candidates in $c$-jets.
\end{itemize}
The three- and four-track candidates are selected in jets that are generator-level-matched to a $b$ or $\bar b$ parton, but they do not form sufficiently clear parent mass peaks to be treated as exclusive $B$ meson reconstruction. We therefore refer to them only as candidate topologies enriched in $B^\pm$ and $B^0/\bar B^0$ decays, respectively, based on their $D^0\pi$ and $D^\pm\pi$ track structures. For each selected candidate, its two, three or four constituent tracks are removed and the remaining jet is rebuilt exactly as it was during training. We then measure
\begin{equation}
 \Delta S = S_{\miss}(X\setminus C)-S_{\miss}(X),
 \label{eq:delta-score}
\end{equation}
where $C$ is the candidate track set. A positive value means that the rebuilt jet looks less complete after the selected tracks are removed. We use two control samples. Mass sidebands outside the signal regions test whether the response is specific to the reconstructed structure. The matched-track control sample removes tracks outside the candidate from the same jet with the same multiplicity and charge pattern. It also matches the removed group as closely as possible in momentum, angular spread and displacement. This tests whether the score increase is more than a generic consequence of deleting charged-particle tracks. Figure~\ref{fig:d0-mass} illustrates the reconstruction for the primary $D^0/\bar D^0\to K^\mp\pi^\pm$ sample in $b$-jets. The peak near the nominal $D^0$ mass defines the signal window, while the two shaded regions provide lower and upper sideband controls. The remaining mass distributions and the mass window definitions are shown in Appendix~\ref{app:candidates}.

The primary comparison is the difference within each jet
\begin{equation}
 \Delta_{\rm pair}=\Delta S_{\rm candidate}-\Delta S_{\rm matched}.
 \label{eq:paired-specificity}
\end{equation}

\subsubsection{Removing reconstructed \texorpdfstring{$D^0$}{D0} candidates from \texorpdfstring{$b$}{b}-jets}

\begin{figure}[tbp]
\centering
\paperplot{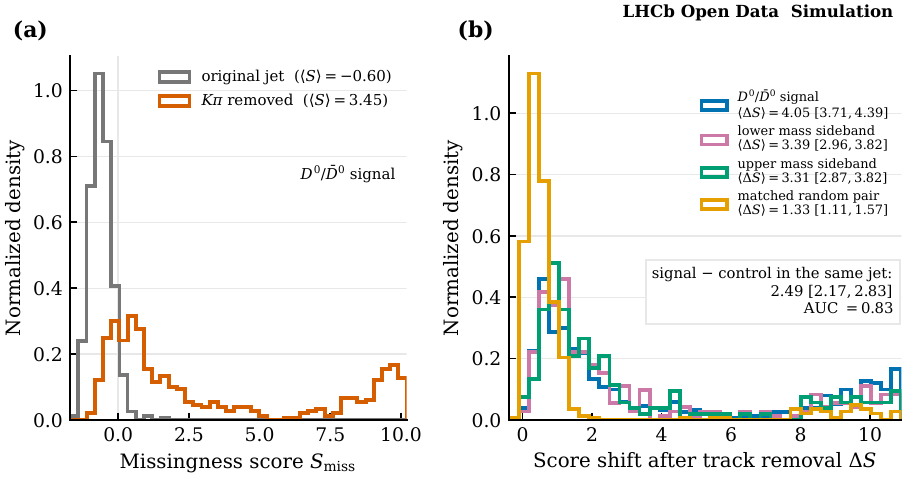}
\caption{Removal of reconstructed $D^0/\bar D^0\to K^\mp\pi^\pm$ candidates from $b$-jets. The left panel compares the score before and after removal, while the right panel shows the score shifts for the signal region, sidebands, and matched tracks.}
\label{fig:d0}
\end{figure}

Figure~\ref{fig:d0} shows the intervention. Removing the $K\pi$ pair from the signal window gives a mean score shift of $4.05$, compared with $1.33$ when a matched pair outside the candidate is removed from the same jet. The mean excess within each jet is $2.49\,[2.17,2.83]$, and it is positive for 88.5\% of the matched pairs. The matched random tracks therefore give a substantially smaller response, showing that the increase is not explained simply by removing two tracks with the same charges and similar kinematics.

The lower and upper sidebands give similar mean shifts of $3.39$ and $3.31$, respectively. The signal window around the $D^0$ peak is higher on average than either sideband, while the agreement between the two sidebands indicates that the response is not driven by one side of the mass distribution. Their distributions nevertheless overlap substantially with the signal distribution. The test therefore does not establish that the model has learned the narrow $D^0$ mass resonance, but it does show sensitivity to removing a coherent, displaced $K\pi$ structure from a $b$-jet.

\FloatBarrier

\subsubsection{Generalization across reconstructed topologies}

\begin{figure}[h]
\centering
\paperplot{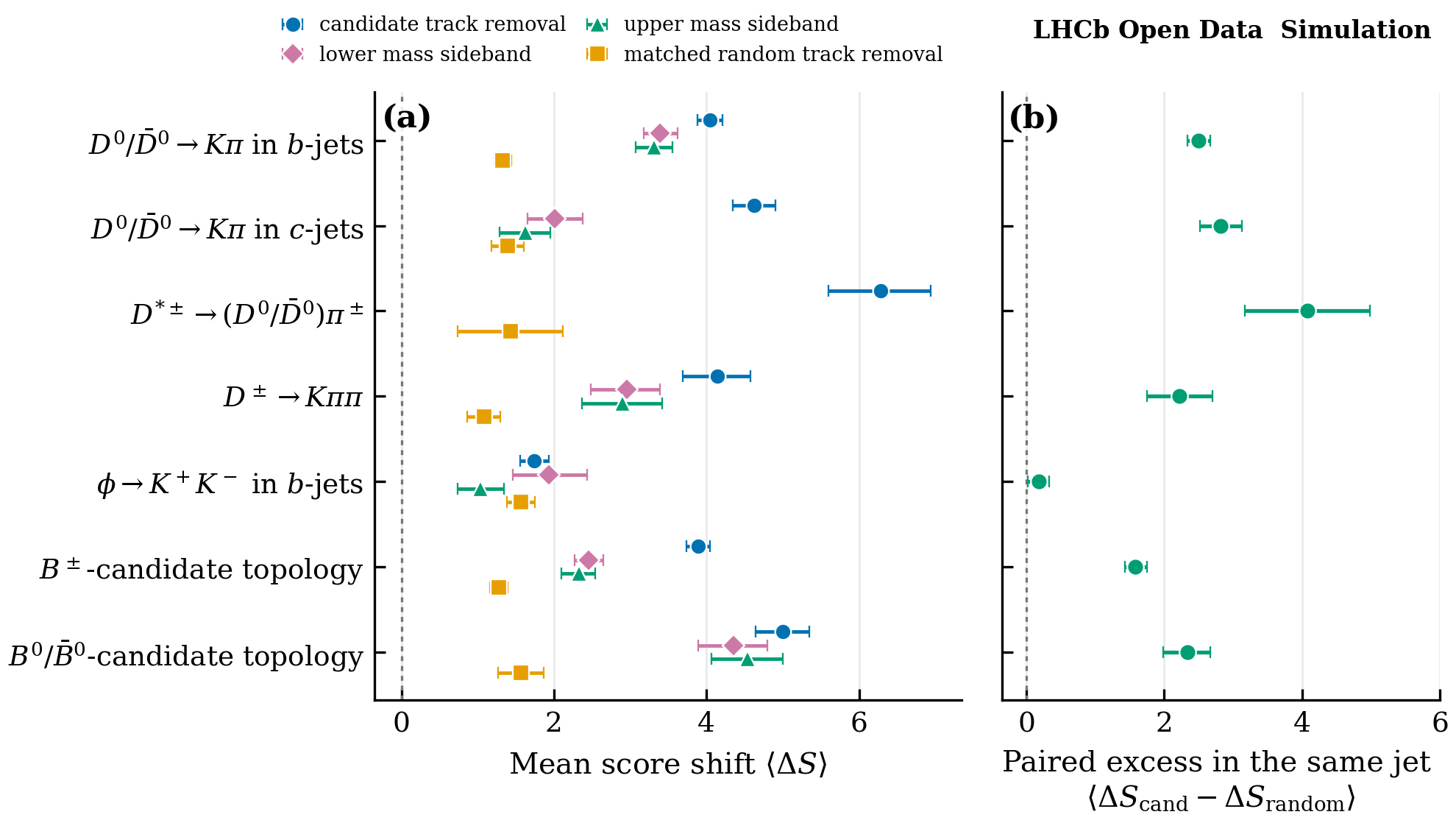}
\caption{Candidate removal tests across reconstructed decay topologies. The left panel compares the score shifts for candidates, sidebands, and matched tracks, while the right panel shows the paired candidate vs. control sample difference.}
\label{fig:decays}
\end{figure}

Having established the response for $D^0/\bar D^0\to K^\mp\pi^\pm$ candidates in $b$-jets, we apply the same test to the other reconstructed topologies. Figure~\ref{fig:decays} compares each candidate removal with a control sample formed from the same number of matched tracks. The sidebands are defined using the mass of the reconstructed resonance being removed. For the two topologies enriched in beauty decays, these are the sidebands of the child $D^0$ or $D^+$ resonance because no parent $B$ mass peak is selected. A larger response to the candidate is observed for $D^0$ decays in both $b$- and $c$-jets, for $D^+$ decays, and for the two topologies enriched in beauty decays. The response also tends to increase with the number of removed particles, as expected from the controlled deletion test in Figure~\ref{fig:synthetic}.

The $D^*$ sample shows the same qualitative candidate vs. control sample difference, but its small matched sample and absence of a usable sideband limit the interpretation. The $\phi\to K^+K^-$ control sample, by contrast, is compatible with no candidate-specific excess. Removing a two particle resonance is therefore not by itself sufficient to produce an enhanced response. Where sidebands are available, their shifts can remain close to those in the signal region, indicating that the model responds to missing visible momentum and correlated displaced structure rather than identifying a narrow mass peak.

Together, these results demonstrate the full methodology in simulation: reconstructed heavy-flavour decays are converted into controlled artificial anomalies and identified by the learned hadronization and decay map across several topologies. This motivates the direct validation in collision data presented next.

\FloatBarrier

\section{Demonstration in LHCb Open Data proton-proton collision samples}
\label{sec:collision-data}

Heavy-flavour production is abundant in the LHCb forward region. At $\sqrt{s}=13$ TeV, the measured production cross sections in the LHCb acceptance are approximately $150\,\mu\mathrm{b}$ for $\bbbar$ and $2800\,\mu\mathrm{b}$ for prompt $\ccbar$ pairs~\cite{LHCb:2016bCrossSection,LHCb:2015charmCrossSection}. The strong angular correlation of the heavy quark pair places approximately one quarter of produced $\bbbar$ pairs within the LHCb acceptance~\cite{LHCb:2014set}. Together with the integrated luminosity collected by LHCb, these rates provide the diverse sample needed to learn a detailed map of heavy hadron hadronization and decay directly from collision data.

To demonstrate this concept, we apply the method to the publicly produced 2017 dijet ntuples. The integrated luminosity recorded in the downloaded files sums to $1.03\,\mathrm{fb}^{-1}$. This sample allows the learned map to be trained and validated entirely in collision data.

\subsection{Collision data sample and independent training}

The sample contains LHCb dijet ntuples for both magnet polarities from CERN Open Data record 94000~\cite{LHCbOpenData:2017dijets}. We select 1,000,000 jets with $p_{\rm T}>25$ GeV and $2.5<\eta<4.0$ for training, together with 200,000 jets in each of the validation, test, and analysis partitions. Jets from the same event are kept in the same partition. The transformer has approximately one million parameters and is initialized randomly and trained only on the collision data training partition. The physical learning task is unchanged and uses the same model architecture.

This collision sample presents a substantially broader learning problem than the heavy-flavour-enriched simulation. It is inclusive with respect to flavour, and no reconstructed secondary vertex is required. Only about 27\% of the jets in the full fiducial sample pass the stored LHCb secondary-vertex tag. The sample is therefore dominated by light jets, with smaller charm and beauty components. Training on this mixture tests whether heavy-flavour hadron structure can be learned without first enriching the data in a particular flavour.

\subsection{Reconstructed control modes}

The data contain exclusive heavy hadron decays that can directly test whether the learned map remains connected to the underlying physics. After training, we reconstruct eight control modes using the particles in each jet. The charm modes are $D^0\to K\pi$, $D^0\to K3\pi$, $D^+\to K\pi\pi$ and $D^{*+}\to D^0\pi^+$. The beauty modes are $B^+\to D^0(K\pi)\pi^+$, $B^+\to D^0(K3\pi)\pi^+$, $B^+\to J/\psi K^+$ and $B^0\to J/\psi K^{*0}$. Together they span different particle multiplicities and decay chains, providing a broad validation of the learned hadronization and decay map.

\begin{figure}[tbp]
\centering
\paperplot{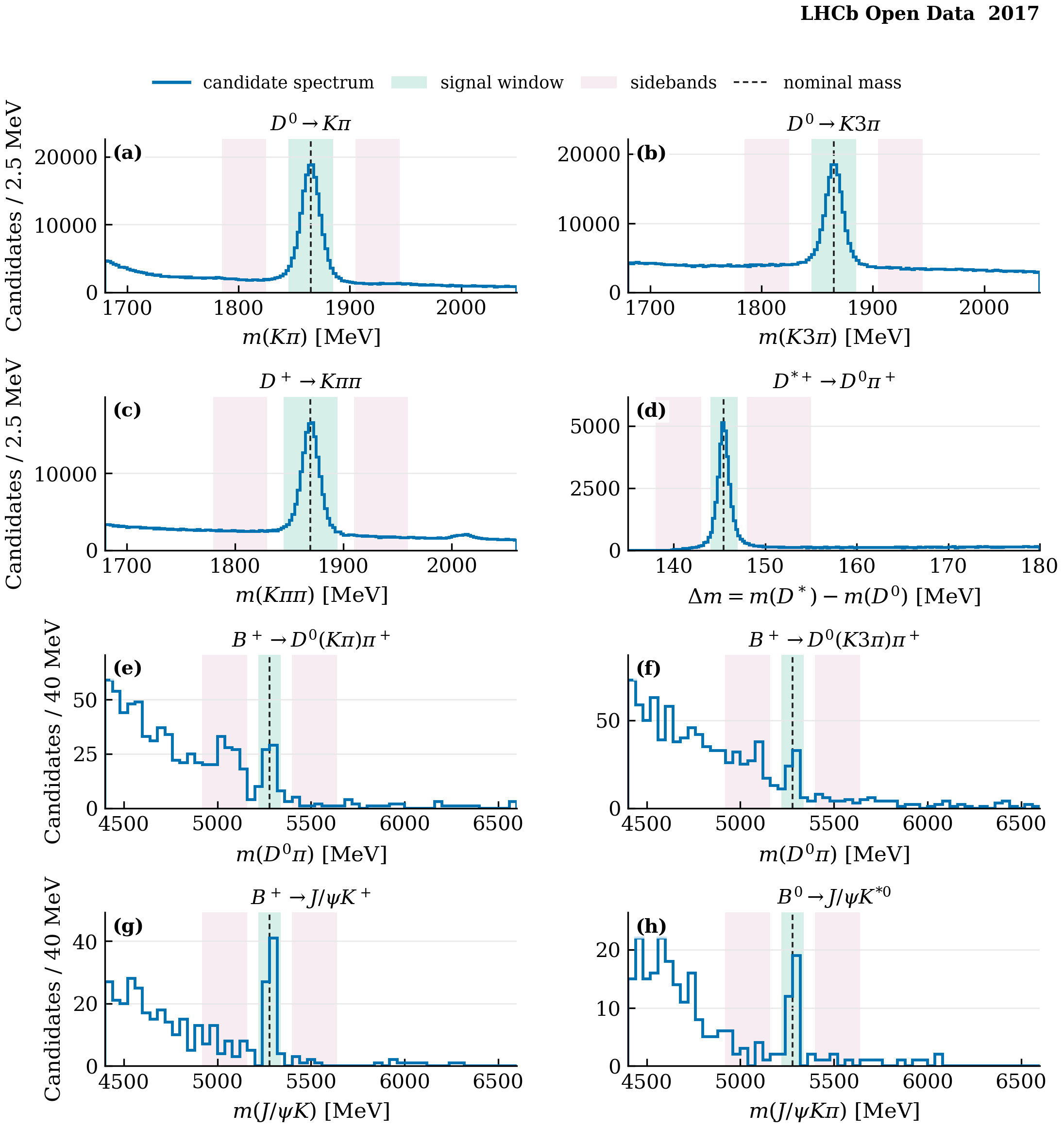}
\caption{Reconstructed charm and beauty control modes in 2017 LHCb collision data. Panels (a)--(d) show charm decays and panels (e)--(h) show beauty decays. Green bands indicate signal windows, purple bands indicate sidebands and dashed lines mark nominal masses. Charge-conjugate modes are combined.}
\label{fig:data-masses}
\end{figure}

Figure~\ref{fig:data-masses} shows the reconstructed candidate spectra across the entire available collision dataset. Signal windows and adjacent sidebands are fixed before evaluating the model on the reserved analysis partition. For the direct charm modes, the sidebands are defined around the reconstructed charm resonance, using $\Delta m$ for the $D^*$ mode. For the beauty modes, the intermediate $D$ or $J/\psi$ is retained in its signal region and the sidebands are defined around the reconstructed $B$ meson candidate.

\subsection{Candidate removal response}

\begin{figure}[h]
\centering
\paperplot{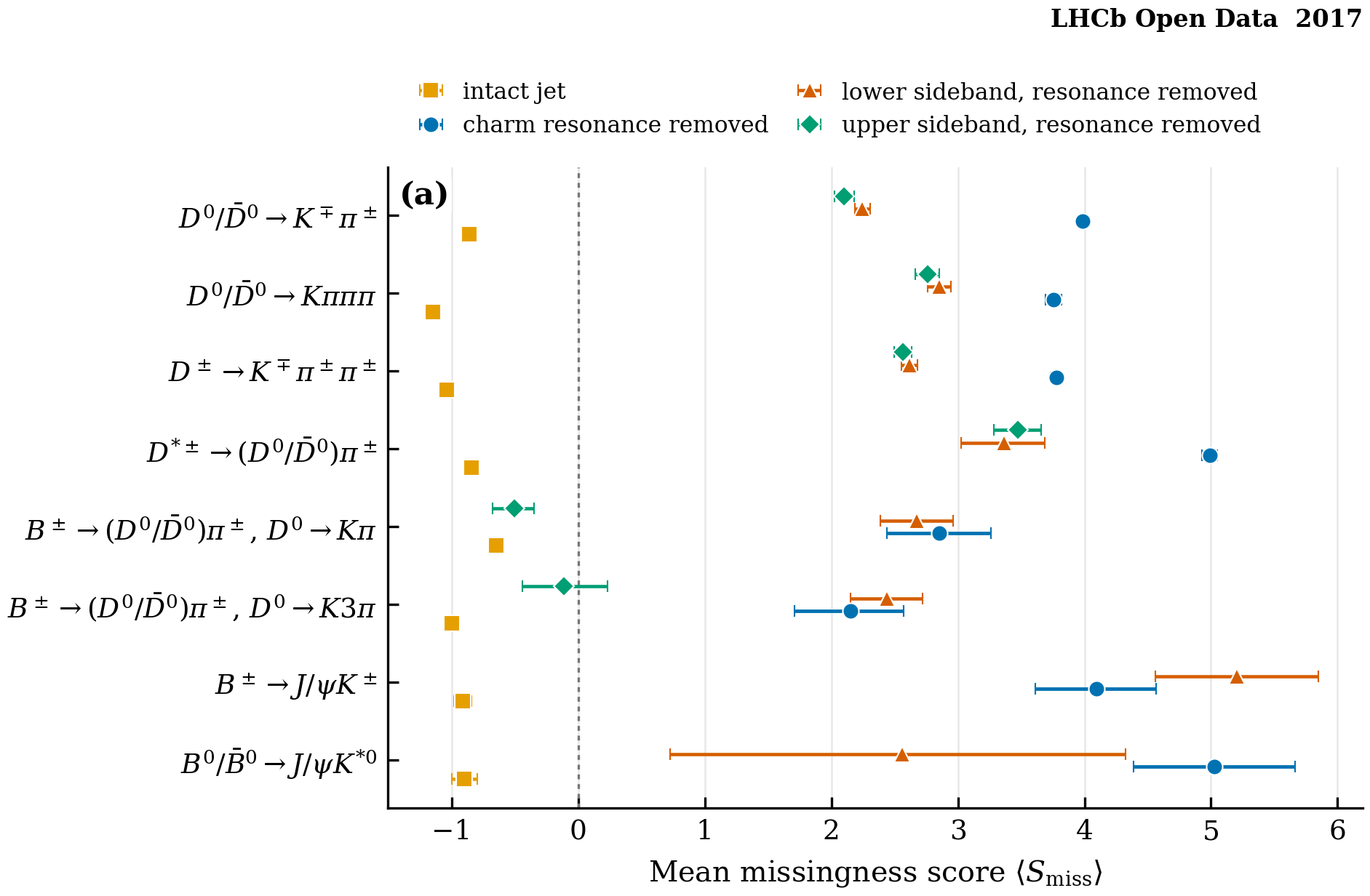}
\caption{Summary of the 2017 collision data demonstration. Squares show intact signal jets, while circles show jets after removing the selected intermediate resonance. Triangles and diamonds show the corresponding lower and upper sideband removals. Horizontal bars show 68\% uncertainty intervals.}
\label{fig:data-score-summary}
\end{figure}

For each signal or sideband candidate, we compare the intact jet with a rebuilt view in which the reconstructed intermediate state is removed. All daughters are deleted for the direct $D^0$ and $D^+$ modes. For $D^*$ and $B\to D\pi$, the reconstructed $D$ is deleted while the bachelor pion is retained. The two muons are deleted for the $J/\psi$ modes. Figure~\ref{fig:data-score-summary} summarizes the resulting mean missingness scores.

In the four charm samples with high statistics, the mean score after removing a candidate in the signal window is 3.99, 3.76, 3.78 and 4.99 for $D^0\to K\pi$, $D^0\to K3\pi$, $D^+\to K\pi\pi$ and $D^*\to D^0\pi$, respectively. Each is larger than both results from adjacent sidebands. The corresponding mean increases within each jet are 4.85, 4.90, 4.82, and 5.83. The response established with controlled simulation is therefore reproduced on reconstructed charm decays in collision data.

The asymmetry between the lower and upper beauty sidebands has a physical origin. The lower mass sidebands are populated by partially reconstructed $B$ meson decays that retain correlated heavy-flavour hadron decay structure. Removing the reconstructed charm or $J/\psi$ resonance can therefore produce an enhanced score, while the upper sidebands are more strongly dominated by combinatorial candidates. The different sideband responses should not be interpreted as a statistical artefact. The four beauty modes contain only 28--66 signal candidates, so the individual comparisons remain qualitative. The collision data result nevertheless shows that the learned completion response acts on reconstructed heavy-flavour decays.

\subsection{Quark charge sign tagging in data with simulation-trained probes}
\label{sec:data-quark-sign}

\begin{figure}[h]
\centering
\paperplot{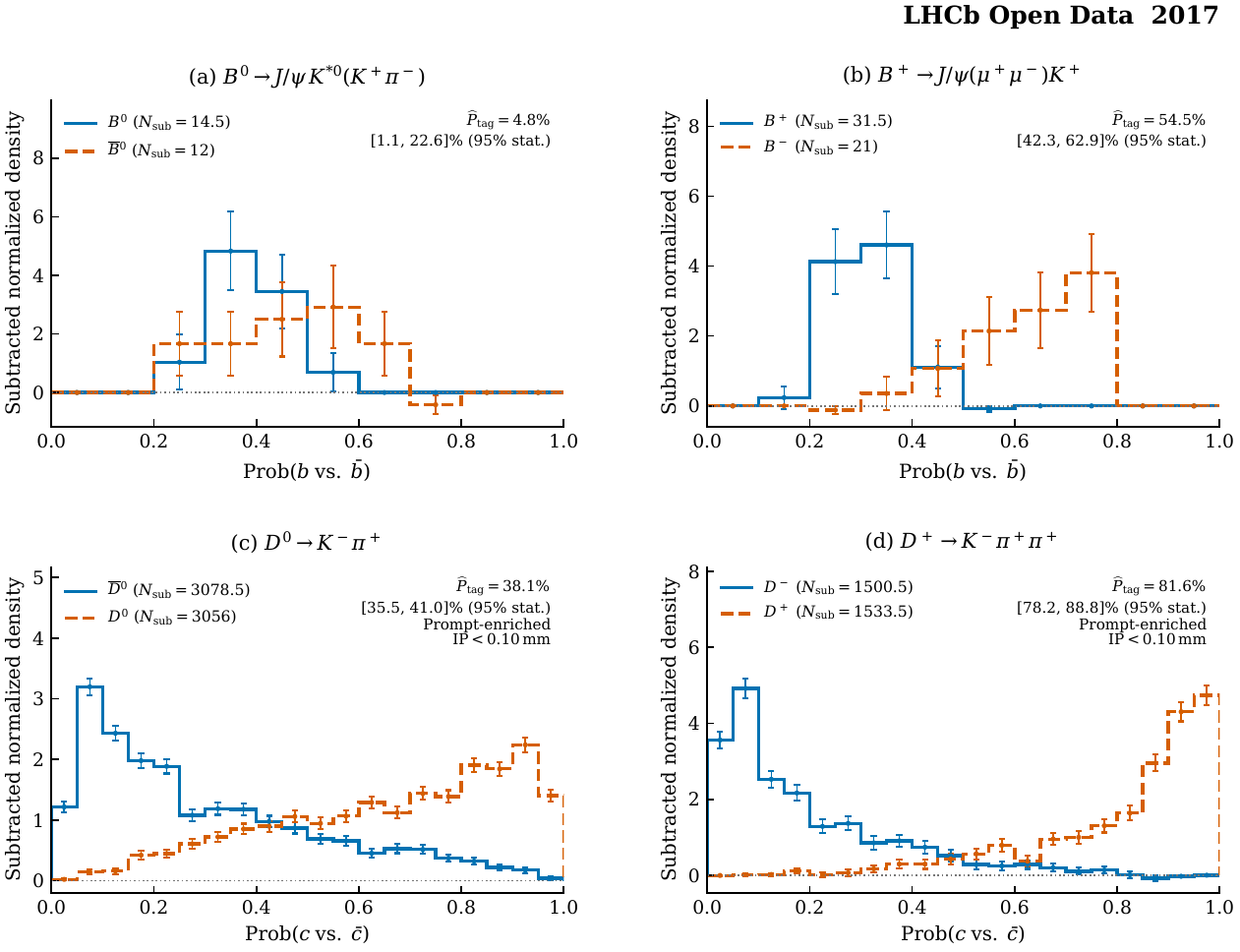}
\caption{Sideband-subtracted distributions of simulation-trained quark charge sign probe outputs for intact jets containing exclusive candidates in 2017 LHCb data: (a) $B^0\to J/\psi K^{*0}(K^+\pi^-)$, (b) $B^+\to J/\psi(\mu^+\mu^-)K^+$, (c) prompt-enriched $D^0\to K^-\pi^+$ and (d) prompt-enriched $D^+\to K^-\pi^+\pi^+$. Charge-conjugate decays are included and normalized separately to unit area after subtraction. $N_{\rm sub}$ denotes the subtracted candidate yield. Larger scores favour $b$ or $c$. Error bars include the statistical uncertainty of the sideband subtraction and normalization. Annotated tagging power estimates use data calibration and have 95\% statistical intervals, conditional on the selected sample and the approximate sideband treatment described in the text.}
\label{fig:data-quark-sign}
\end{figure}

Exclusive reconstructed decays provide a direct test of whether the flavour information learned in simulation survives in real collision data. We apply the encoder and affine quark charge sign probes from Section~\ref{sec:flavour-probes} without changing their simulation-trained parameters. This is distinct from the independently trained collision data model used above. Inference uses the intact jet containing the reconstructed candidate, with no candidate constituents removed. Figure~\ref{fig:data-quark-sign} shows the $b$ versus $\bar b$ scores for $B^0\to J/\psi K^{*0}$ and $B^+\to J/\psi K^+$ and the $c$ versus $\bar c$ scores for $D^0\to K^-\pi^+$ and $D^+\to K^-\pi^+\pi^+$, including charge-conjugate decays. The reconstructed $D^0$ and $D^+$ are selected to be prompt by requiring a geometric candidate impact parameter below $0.10$~mm. The reconstructed final-state charges provide the reference decay-flavour labels. For neutral candidates these are not production-flavour labels, and no mixing correction is applied. For $B^0\to J/\psi K^{*0}$, the kaon charge in $K^{*0}\to K^+\pi^-$ identifies the decay flavour.

We estimate the combinatorial background contribution using the fixed lower and upper mass sidebands. For each charge category, their summed score histogram is scaled by the ratio of signal window to combined sideband widths and subtracted before normalization to unit area. This factor is $1/4$ for the beauty modes and $1/2$ for the charm modes. The retained sideband sample excludes events containing a signal-window candidate and uses a separate candidate choice in each region.

We calculate the tagging power using eq.~\eqref{eq:tagging-power}, with the same subtraction weights. The resulting estimates are $54.5\%$ for $B^+\to J/\psi K^+$, $38.1\%$ for prompt-decay-enriched $D^0\to K\pi$ and $81.6\%$ for prompt-enriched $D^+\to K\pi\pi$. For $B^0\to J/\psi K^{*0}$, the estimate is $4.8\%$, with a broad 95\% statistical interval of $[1.1,22.6]\%$ from only 28 signal-window candidates.

The pronounced charge separation for charged beauty and the two charm modes provides a data verification of useful flavour information learned in simulation. None of these collision events was used to train the affine probes obtained from simulation. The observed separation thus demonstrates transfer to collision data. This result extends the evidence beyond simulation benchmarks to identifiable heavy-flavour decays with data.

\subsection{Exploratory similarity structure of anomalous jets}
\label{sec:data-anomaly-similarity}

\begin{figure}[tbp]
\centering
\paperplot{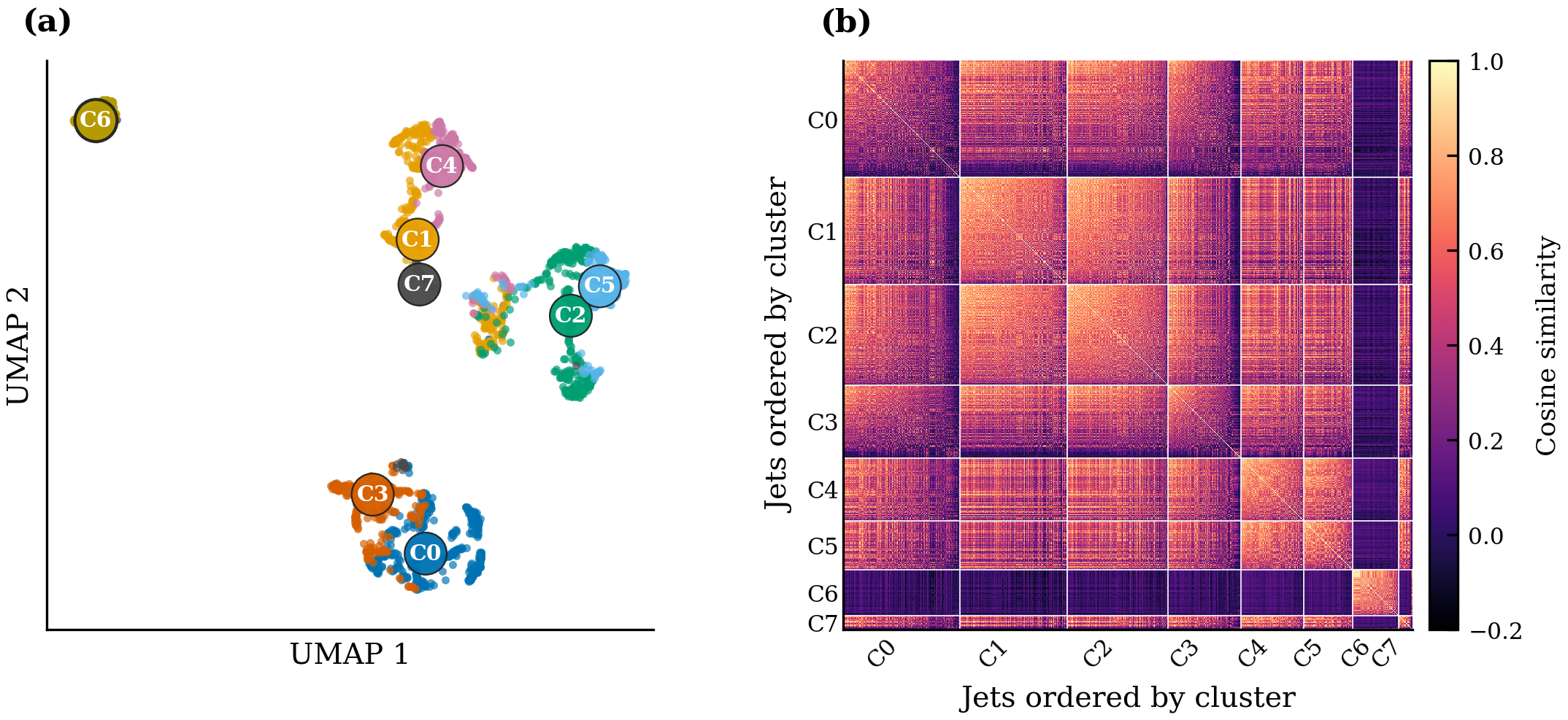}
\caption{Exploratory structure of the 2000 highest-scoring collision-data jets in the restricted fiducial region, taking the highest-scoring 1000 jets from each magnet polarity. (a) UMAP projection with the descriptive group labels. Labels C0--C7 denote the descriptive eight-group partition. (b) Pairwise jet--jet cosine similarity, with jets ordered by group and then by similarity to the group centroid. White lines mark group boundaries and the diagonal entries are unity.}
\label{fig:data-anomaly-map}
\end{figure}

Next, we perform the inference on the full 2017 collision data and calculate the missingness score of each jet. As an exploratory application of the learned representation, we then restrict the sample to jets with $p_{\rm T}>30$~GeV and $2.5<\eta<4.0$, and retain 1000 jets with the largest missingness score from each magnet polarity, giving 2000 jets in total. For each pair of intact jets we calculate the cosine similarity of their learned representations. Figure~\ref{fig:data-anomaly-map} shows a two-dimensional Uniform Manifold Approximation and Projection (UMAP) visualization~\cite{McInnes:2018}, in which nearby points represent jets with similar reconstructed content, together with the full jet-jet similarity matrix. For visual organization, the jets are partitioned into eight groups, and the matrix is ordered first by group and then by similarity to the corresponding centroid. The choice of eight is a descriptive resolution for this survey. It is not evidence for eight distinct physical populations.

Table~\ref{tab:data-anomaly-clusters} gives compact, descriptive labels based on the reconstructed constituent content of each group. These labels summarize relative enrichment within this selected sample and do not identify an underlying process. Several groups are dominated by neutral constituents, whereas C6 is visibly different: its 163 jets contain a mean of 11.65 charged and 0.10 neutral constituents. The median charged multiplicity is 12, with a range from 4 to 20, and 146 of the 163 jets have no stored neutral constituents.

\begin{table}[tbp]
\centering
\caption{Descriptive summary of the eight groups in the high-score similarity survey. The final two columns are mean reconstructed constituent counts per jet. The labels describe relative content and carry no process interpretation.}
\label{tab:data-anomaly-clusters}
\begin{tabular}{@{}clrrr@{}}
\toprule
Group & Descriptive content & Jets & $\langle N_{\rm charged}\rangle$ & $\langle N_{\rm neutral}\rangle$ \\
\midrule
C0 & photon and $\Lambda$ enriched              & 410 &  2.27 & 6.51 \\
C1 & neutral rich, positive-charge enriched      & 376 &  1.59 & 8.83 \\
C2 & neutral rich, negative-charge enriched      & 354 &  1.32 & 8.21 \\
C3 & photon and $\Lambda$ enriched, positive charge & 257 & 2.31 & 6.48 \\
C4 & $K^0_{\rm S}$ enriched, positive charge    & 220 &  1.15 & 4.83 \\
C5 & $K^0_{\rm S}$ enriched, negative charge    & 171 &  1.26 & 4.31 \\
C6 & charged rich, neutral poor                  & 163 & 11.65 & 0.10 \\
C7 & mixed content, higher $p_{\rm T}$          &  49 &  1.92 & 4.14 \\
\bottomrule
\end{tabular}
\end{table}

\begin{figure}[!t]
\centering
\paperplot[0.93\textwidth]{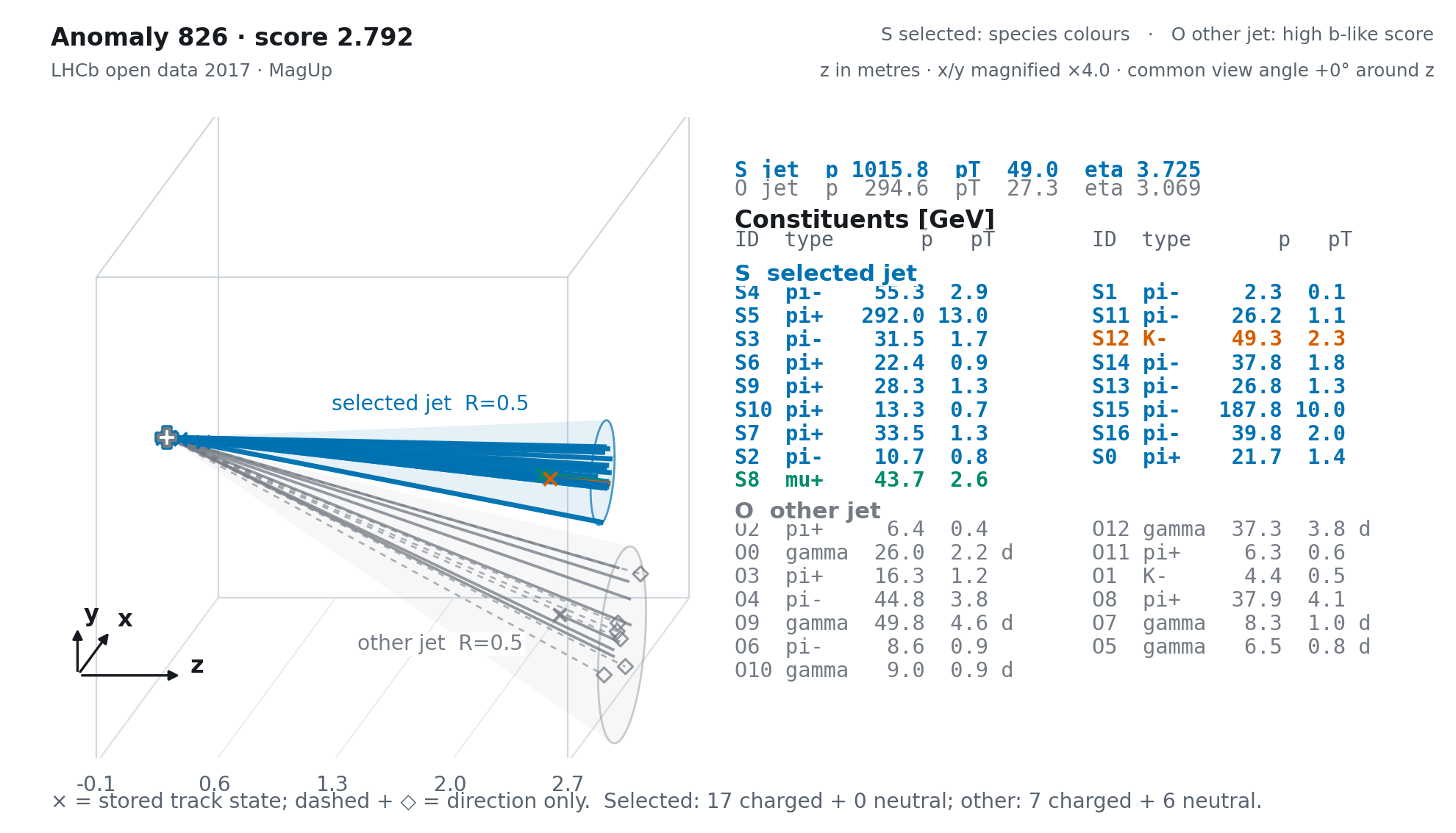}\\[-1mm]
\paperplot[0.93\textwidth]{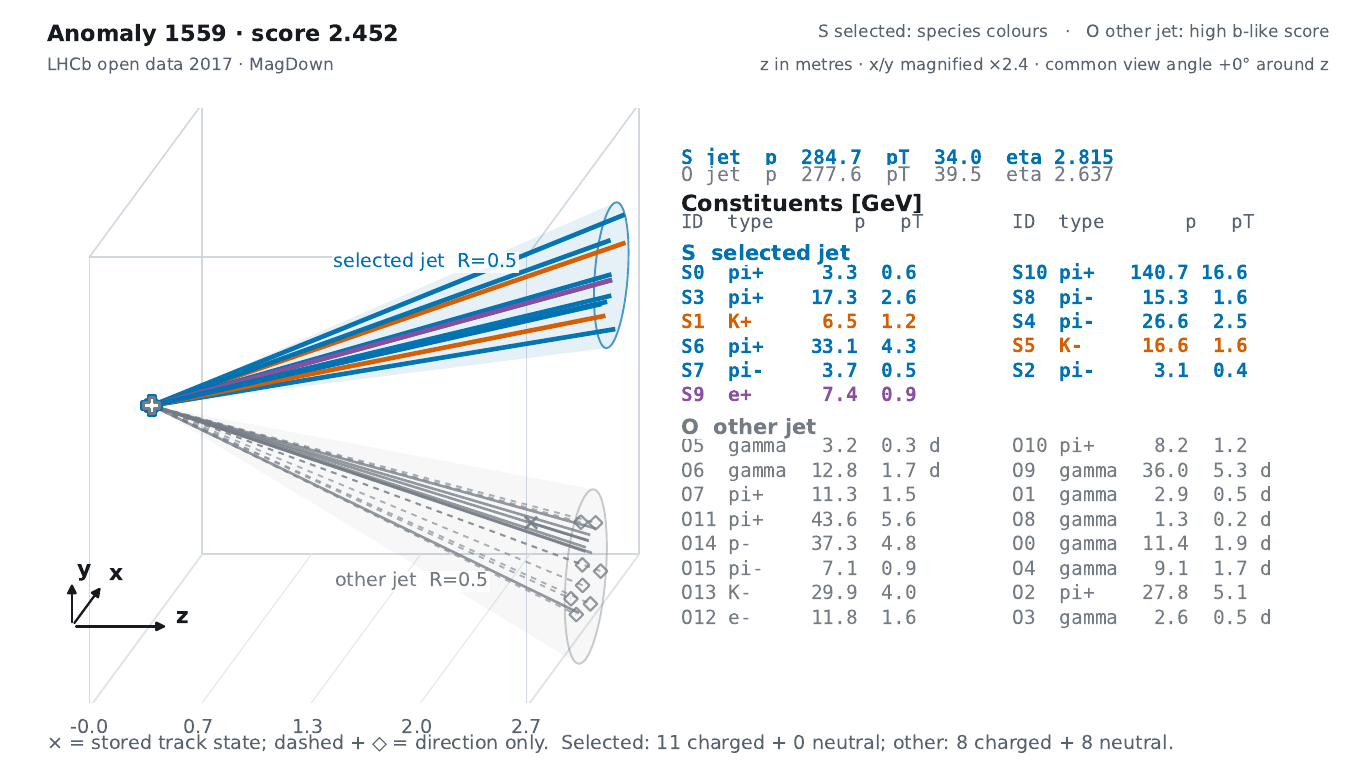}
\caption{Reduced event displays for two charged-rich, neutral-poor jets in C6: rank 826 (upper) and rank 1559 (lower). The selected jet is shown in particle species colours and the other jet in grey. Solid charged-particle trajectories start at their stored reference states. Dashed trajectories show direction-only constituents, and the lightly shaded regions indicate the projected $R=0.5$ jet cones. The transverse directions are magnified relative to the beam direction. These displays contain neither detector hits nor magnetic field propagation and do not determine the origin of the anomalous score.}
\label{fig:data-anomaly-events}
\end{figure}

We inspect C6 because its charged-rich, neutral-poor content is unusual within this high-score sample. As a recurrence check, we search all 9,858,223 jets passing the same fiducial selection for similarity to C6. Requiring zero stored neutral constituents, a negative missingness score and a maximum cosine similarity of at least 0.9 to any C6 jet gives 1630 matches outside the high-score tail. Their mean charged multiplicity is 7.63, with a median of 7, compared with a mean of 11.65 and a median of 12 in C6. The charged-rich, neutral-poor pattern therefore also occurs at low anomaly score, while C6 occupies a higher-multiplicity part of this population. To illustrate the topology, Fig.~\ref{fig:data-anomaly-events} shows two C6 events for which the other jet passes the stored secondary-vertex tag and lies in a beauty-like region of the legacy tagger response. Rank 826 has 17 charged and zero neutral selected-jet constituents, while rank 1559 has 11 charged and zero neutral constituents and provides especially clear separation between the two jets. This event choice is made for visual clarity and is not a statistical selection.

This study is therefore presented only as an example of how cosine-similarity retrieval can organize high-scoring events and expose a population for closer inspection. The lack of stored neutral constituents refers to the reduced jet record and, by itself, does not distinguish a physical topology from a detector or reconstruction effect. A dedicated follow-up would require a larger model, more collision data, stability tests across data-taking conditions and access to the complete LHCb event record, including detector hits and calibrated track and neutral reconstruction. Such a study could use the same similarity-based workflow to determine whether this population remains coherent and to establish its origin. With larger datasets and access to the complete event record, the anomaly detection method illustrated here, coupled with opposite-side tagging and the jet-based flavour and charge taggers developed in this study, could provide a data-driven strategy to probe $b$- and $c$-hadron decays for signs of physics beyond the SM.

\FloatBarrier

\section{Discussion}
\label{sec:discussion}

This study provides a proof of principle that the precise reconstruction of LHCb can be used to learn a map of heavy-flavour hadronization and decay. The evidence comes from several complementary results. The missingness score increases systematically as particles are removed, showing that the transformer learns the completion task on unseen events. Affine probes of the learned representation outperform the same transformer with random weights and recover 97--100\% of the fully supervised performance across five flavour and quark charge sign classification tasks. Removing reconstructed decay products in simulation produces a larger response than removing matched tracks, connecting the learned map to correlated heavy-flavour hadron decay structure. Finally, a model trained independently on inclusive collision data reproduces the expected response in reconstructed charm decays and extends the test to exclusive beauty decays. Together these results demonstrate the map without relying on a single decay hypothesis or supervised flavour labels.

The LHCb experiment is especially well suited to this construction. Its precise vertexing resolves displaced decay topologies, while the RICH and calorimeter systems provide complementary particle identification. The abundant heavy-flavour sample supplies the diversity needed for self-supervised training. Charge and magnet polarity are retained separately, allowing the network to represent the detector configuration. Reversing the magnet changes charge-dependent detector effects while leaving the underlying decay physics unchanged. A semi-invisible decay signature should therefore remain consistent between polarities, while a detector-induced anomaly may change. Magnet reversal can thus help distinguish physical signatures from detector effects and, with the necessary calibration, could enable comparisons of charge-conjugate decays in measurements of $CP$ violation.

In an unrestricted collision sample, the largest initial anomaly populations are expected to arise from detector effects, reconstruction failures and the boundaries of the LHCb acceptance. Requirements such as $p_{\rm T}>25$ GeV and $2.5<\eta<4.0$ should therefore be considered to reduce contributions from the jet momentum threshold region and the edges of the detector acceptance. They do not replace dedicated checks of detector conditions, periods of data taking and magnet polarity, which would be required before interpreting anomalous populations in a physics analysis.

Jets provide a compact environment for this first demonstration, and the present study treats each jet independently. The learning objective does not depend on jet clustering and should therefore generalize to heavy-flavour hadron decays reconstructed without a jet requirement, similar to the newly developed CMS same-sign tagger operating on charged particles within an $R = 0.8$ cone~\cite{CMS-PAS-BPH-26-005}. A natural LHCb extension would represent the usage of jets clustered with a larger distance parameter. One could go even further by utilizing the entire event by using reconstructed particles centred on a decay candidate or the primary vertex. Such a representation could use the full fragmentation information together with the opposite-side and same-side information employed by the existing flavour tagging methods. Controlled signal injection, larger reconstructed control samples and validation resolved by polarity would then allow the map to be calibrated for a specified physics analysis.

The learned representation also enables several extensions beyond the missingness score. Information retrieval based on cosine similarity could identify jets, and eventually complete events, with related reconstructed content. For an anomalous event, comparison with nearby events that have ordinary scores could help determine whether the response is associated with a known decay topology or a detector effect. With a larger model trained on more collision data, calibrated completion predictions could also help characterize anomalous events by identifying plausible missing particles and their properties. These predictions could be combined into observables sensitive to rare reconstructed decay topologies, providing candidates for dedicated searches for rare SM processes.

\section{Conclusions}
\label{sec:conclusion}

We have shown that the precise reconstruction of LHCb can be used to learn a self-supervised map of heavy-flavour hadronization and decay. Without either flavour or exclusive decay labels, the self-supervised transformer model learns to reconstruct masked particle identification labels and removed particles. In simulation, the learned representation outperforms the same transformer with random weights and recovers 97--100\% of fully supervised performance across five flavour and quark charge sign tasks. Controlled particle removals and reconstructed decay interventions both produce the expected ``missingness'' response.

Crucially, the simulation-trained encoder and affine probes also separate $B^+$ and charm decay flavour categories in LHCb Open Data of 2017 proton-proton collisions at $\sqrt{s} = 13$~TeV that were never used in their training. These exclusive reconstructions test the learned flavour information in a realistic detector.

The same method, trained independently on one million 2017 collision data jets, responds systematically to eight reconstructed charm and beauty decays. For the four high-statistics charm modes, the signal response exceeds that in both adjacent sidebands. These results establish a proof of principle that collision data can define a learned map of heavy hadron decays. Extended to complete events and calibrated across magnet polarities, such a map could open searches for incomplete and unusual decays.

\appendix
\raggedbottom

\section{Training diagnostics}
\label{app:training}

\begin{figure}[H]
\centering
\paperplot{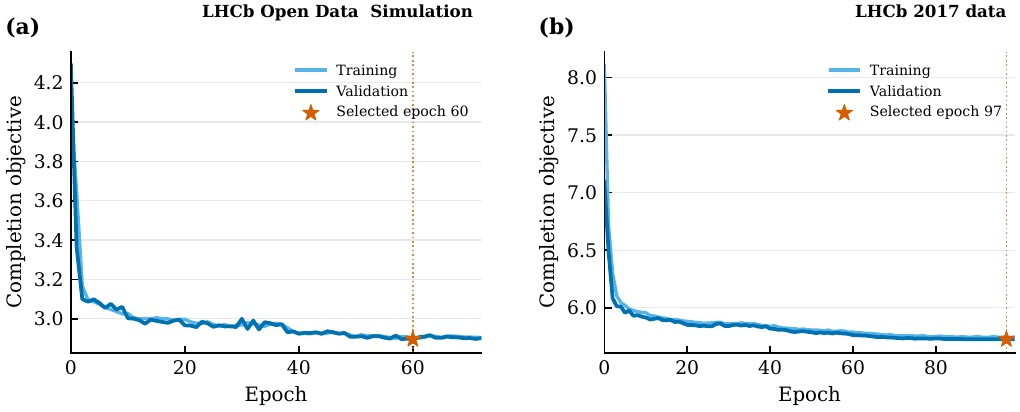}
\caption{Completion objective during training on simulation (a) and 2017 collision data (b). Stars mark the selected checkpoints.}
\label{fig:training-history}
\end{figure}

Figure~\ref{fig:training-history} shows the completion objective for the models trained on simulation and collision data. The selected checkpoints are epochs 60 and 97. Their training and validation objectives are 2.90 and 2.90 in simulation, and 5.75 and 5.73 in data. Table~\ref{tab:head-metrics} summarizes the corresponding completion performance on the held-out samples.

\begin{table}[H]
\centering
\small
\begin{tabular}{lcrr}
\toprule
Task & Metric & Simulation & 2017 data \\
\midrule
Missing count & Accuracy & 0.396 & 0.354 \\
Missing particle existence & ROC AUC & 0.70 & 0.75 \\
Missing charge & Equal class accuracy & 0.830 & 0.751 \\
Balanced missing PID & Equal class accuracy & 0.461 & 0.352 \\
\bottomrule
\end{tabular}
\caption{Held-out performance of the principal completion heads. The simulated and collision data samples contain different reconstructed particle mixtures and are not intended as a direct benchmark against one another.}
\label{tab:head-metrics}
\end{table}

\begin{figure}[H]
\centering
\paperplot{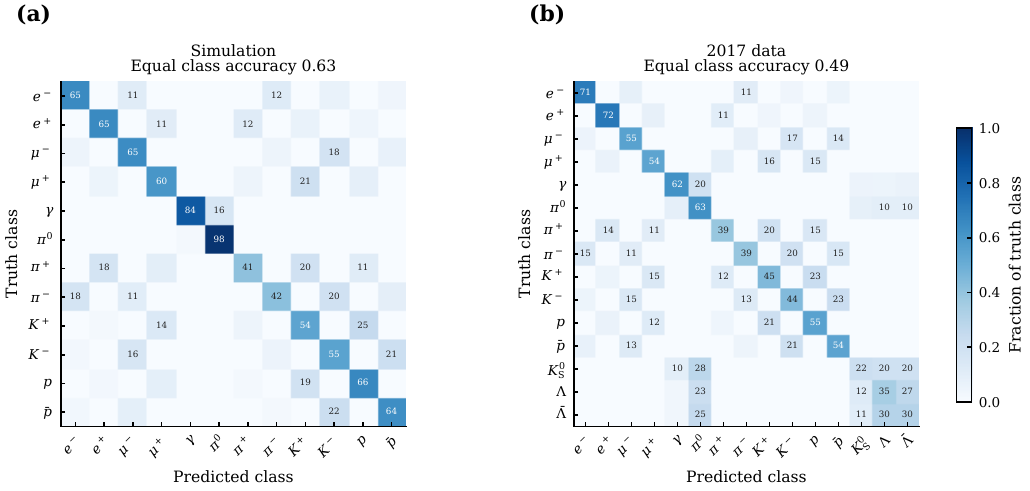}
\caption{Row-normalised confusion matrices for PID prediction balanced across classes for visible particles in simulation (a) and 2017 collision data (b).}
\label{fig:validation-heads}
\end{figure}

Figure~\ref{fig:validation-heads} shows the PID prediction balanced across classes for visible particles whose PID inputs are masked. The equal class accuracy is 0.63
in simulation and 0.49 in collision data.

\FloatBarrier
\section{Reconstructed candidate definitions}
\label{app:candidates}

Candidates are reconstructed from charge, PID, displacement and track geometry, independently of the missingness score. Table~\ref{tab:candidate-selections} records the frozen signal and sideband definitions. All charge-conjugate modes are included, and one candidate per jet is selected without using the model response.

\begin{table}[H]
\centering
\scriptsize
\setlength{\tabcolsep}{5pt}
\begin{tabular}{p{0.14\textwidth}p{0.12\textwidth}p{0.31\textwidth}p{0.32\textwidth}}
\toprule
Category & Topology & Signal region & Sideband or control \\
\midrule
$D^0$ in $c$-jet & $K^\mp\pi^\pm$
& $1840<m(K\pi)<1890$ MeV
& $1790$--$1840$ or $1890$--$1940$ MeV \\
\addlinespace
$D^{*\pm}$ & $D^0\pi^\pm$
& $144<\Delta m<147$ MeV and $|m(K\pi)-m_{D^0}|<25$ MeV
& $140$--$143$ or $148$--$151$ MeV in $\Delta m$ \\
\addlinespace
$D^\pm$ & $K^\mp\pi^\pm\pi^\pm$
& $1850<m(K\pi\pi)<1890$ MeV
& $1810$--$1850$ or $1890$--$1930$ MeV \\
\addlinespace
$\phi$ in $b$-jet & $K^+K^-$
& $1005<m(KK)<1035$ MeV
& $975$--$1005$ or $1035$--$1065$ MeV \\
\addlinespace
$B^\pm$ enriched & $D^0\pi^\pm$
& $D^0$ signal window
& $D^0$ sideband or three-track control \\
\addlinespace
$B^0/\bar B^0$ enriched & $D^\pm\pi^\mp$
& $D^\pm$ signal window
& $D^\pm$ sideband or four-track control \\
\bottomrule
\end{tabular}
\caption{Frozen mass windows for reconstructed candidates in the simulation removal tests. The primary $D^0$ study in $b$-jets uses $|m(K\pi)-m_{D^0}|<20$ MeV and $40<|m(K\pi)-m_{D^0}|<80$ MeV for its sideband. The beauty enriched categories use the sideband of the reconstructed charm decay.}
\label{tab:candidate-selections}
\end{table}

\begin{figure}[H]
\centering
\paperplot{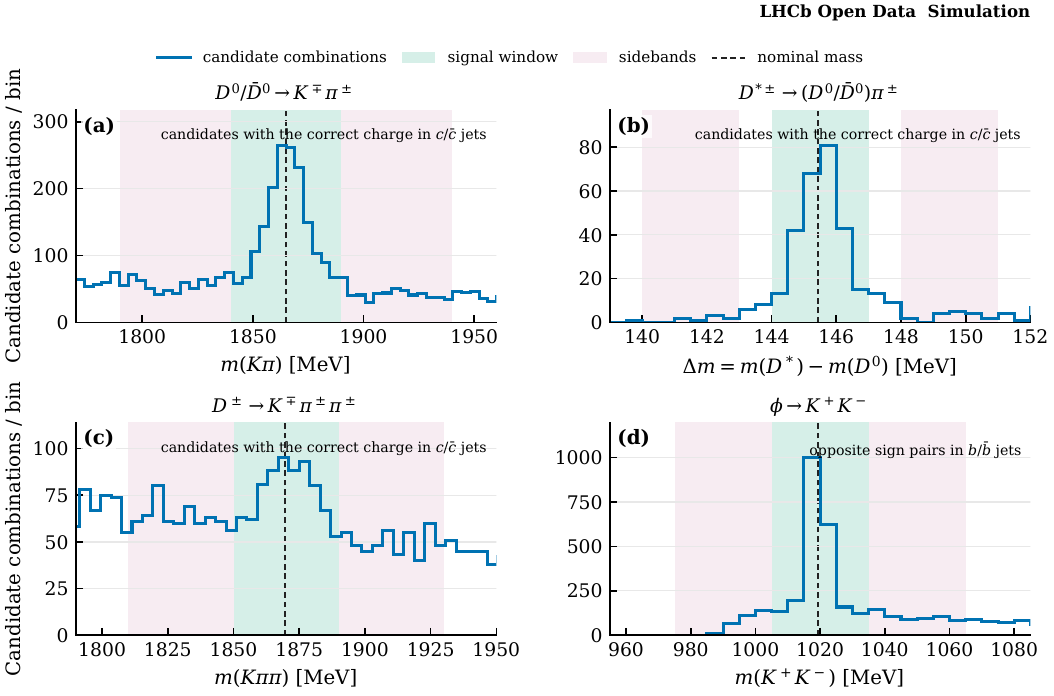}
\caption{Reconstructed candidate masses used to define the signal windows and adjacent sidebands for the simulation removal tests.}
\label{fig:candidate-masses}
\end{figure}

\acknowledgments

We thank the LHCb Collaboration and CERN Open Data Portal for making the simulated jet samples publicly available. We thank Lesya Shchutska for the useful discussion. This work utilized computing resources at Brown University's Center for Computation and Visualization and is partially supported by the Office of High Energy Physics of the U.S. Department of Energy under the contract DE-SC0010010. The authors used generative AI tools during the development of this work for coding assistance and language editing. The scientific content, interpretation, and conclusions were developed entirely by the authors.

\bibliographystyle{JHEP}
\bibliography{biblio}

\end{document}